\documentclass[conference]{IEEEtran}

\usepackage{graphics} 
\usepackage[font=footnotesize]{caption}
\usepackage[config=altsf]{subfig}
\usepackage{epsfig}
\usepackage{cite}
\usepackage{amsmath,amssymb,amsfonts}
\usepackage[table,xcdraw]{xcolor}
\usepackage{algorithmic}
\usepackage{graphicx}
\usepackage{textcomp}
\usepackage{xcolor}
\usepackage{svg}
\usepackage{lettrine}
\usepackage{threeparttable,booktabs}
\usepackage{multirow}
\usepackage[multiple]{footmisc}
\usepackage{hyperref}

\definecolor{blue_fill}{RGB}{215,227,240}
\definecolor{red_fill}{RGB}{255,213,213}
\definecolor{green_fill}{RGB}{160,233,193}
\definecolor{violet_fill}{RGB}{218,134,173}

\newcommand{\celdaBlue}{\cellcolor{blue_fill}}

\makeatletter
\def\ps@IEEEtitlepagestyle{%
  \def\@oddfoot{\mycopyrightnotice}%
  \def\@evenfoot{}%
}
\def\mycopyrightnotice{%
  {\footnotesize This work has been submitted to the IEEE for possible publication. Copyright may be transferred without notice, after which this version may no longer be accessible.\hfill}
  \gdef\mycopyrightnotice{}
}

\begin{document}

\bstctlcite{IEEEexample:BSTcontrol}

\title{All-Digital FPGA-based DAC with None or Few External Components}

\author{\IEEEauthorblockN{Luis E. Rueda G., Edward Silva, Andres Centeno, and Elkim Roa\\ Integrated Systems Research Group - OnChip, Universidad Industrial de Santander \\ Bucaramanga - Colombia \\ Email:\ \{leruegue,efroa\}@uis.edu.co}}

\IEEEoverridecommandlockouts

\maketitle

\begin{abstract}
One of the many limitations with the mixed-signal design is physically testing circuit ideas. While it is easier to test digital circuits with FPGAs, this can not be done usually with mixed-signal circuits. Although some FPGAs have built-in analog-to-digital and digital-to-analog converters, regular commercial FPGAs development boards and low-cost FPGAs lack built-in data converters. Here we introduce an all-digital FPGA-based DAC, which is one of the main blocks to enable mixed-signal experiments. The DAC can be synthesized entirely in an FPGA and does not require the use of external components. Furthermore, and to extend its range of applications, a discussion regarding the proposed DAC's problems and possible solutions is presented. Experimental demonstration of a 4-bit and a 5-bit DAC corroborate the theoretical analysis developed in this work. This work also suggests a scheme which includes few external resistors to improve the linearity (DNL$\leq$0.25LSB and an INL$\leq$0.5LSB), and the power consumption (5X improvement over the standalone configuration).
\end{abstract}


\IEEEpeerreviewmaketitle

\section{Introduction}

One of the main disadvantages of mixed-signal circuits, in comparison with all-digital systems, is the complex design process involved and the many different ways to implement this design. In particular, there is not a simple way to physically construct a proof-of-concept of a mixed-signal system. On the digital counterpart, this can be done easily by synthesizing the digital description in an FPGA. To test mixed-signal circuits ideas and implementations, it is necessary to either construct a discrete circuitry, possibly with an FPGA for the digital part and commercially available integrated circuits (IC) or to fabricate an IC, which is neither a simple nor a cheap process.

Many of the signal processing within mixed-signal systems is done in the digital domain, and use analog-to-digital (ADC) and digital-to-analog (DAC) converters to interact with the analog part of the system. Some FPGAs already include internal ADCs and DACs \cite{Xilinx_fpga}. These FPGAs might be used to test mixed-signal circuits directly. The problem is that standard and low-cost FPGAs do not have internal converters \cite{icoboard}, hence limiting their use to digital implementations only. There are also field-programmable analog arrays (FPAAs) \cite{Anadigm_fpaa}, as well as programmable system-on-chip (PSoC) \cite{Cypress_psoc} in the market, which would be the analog counterpart of the FPGAs, but their elevated prices (mainly the FPAAs), and their lack of versatility, make these solutions unsuitable for many applications. With the latter in mind, this work is focused on the implementation of an FPGA-based digital-to-analog converter that might be described using an HDL language for synthesis.

There are several ways to implement an FPGA-based DAC: single-bit stream DACs (e.g., PWM and $\Delta\Sigma$) \cite{Magrath1997}, and multi-bit DACs (e.g., binary-weighted DAC, R-2R DAC) \cite{Zet2019}. To the authors' knowledge, all of the existing FPGA-based DAC implementations need external components. In fact, all multi-bit DACs base their operation on a mixed-signal approach where the FPGA implementation is the control unit used to switch on or off external components. The single-bit counterparts are closer to be called a true FPGA-based DAC implementation, but still, they need external components to filter high-frequency harmonics in order to obtain a proper digital-to-analog conversion \cite{Lonla2015}. Furthermore, the single-bit implementations usually limit their application to low-frequency operation (as a difference to the multi-bit approach).

This work proposes an FPGA-based DAC completely synthesized and instantiated in a generic FPGA without requiring additional external components. The proposed DAC is a multi-bit implementation whose sampling rate is mainly limited by the FPGA's GPIO (general purpose input/output) dynamic characteristics. The latter allows its usage in a broader frequency range than other FPGA-based DACs. Furthermore, to expand the DAC's application to other uses, the problems and non-idealities of the proposed DAC, as well as possible solutions involving the inclusion of few external resistors, are discussed. 

\section{Proposed FPGA-based DAC}\label{sec:fpga_dac}

\begin{figure}[!t]
	\centering
	\includegraphics[width=1.0\linewidth]{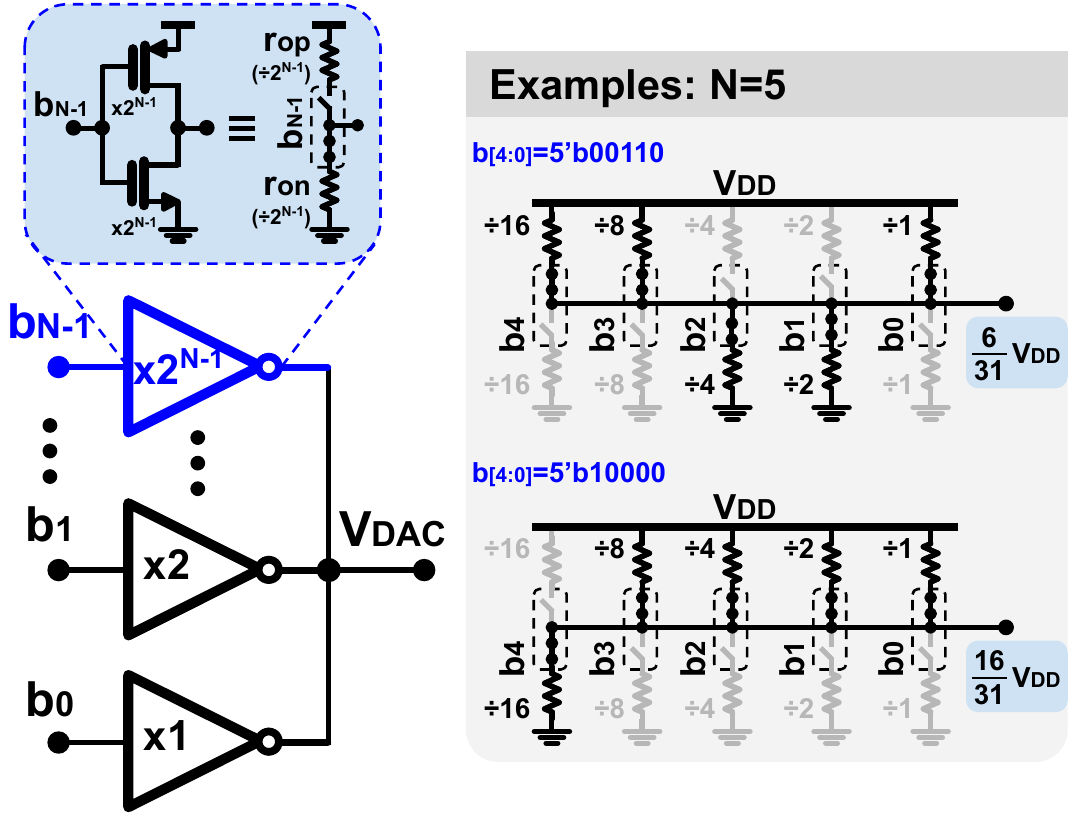}
    \caption{\small Inverter-based DAC principle. In an ideal case, the DAC behaves like a resistor-DAC voltage divider.}
    \label{fig:dac_principle}
\end{figure}

The FPGA-based DAC is an application of the inverter-based DAC principle \cite{Deng2015}. A voltage DAC can be developed by shorting the output of several inverters, as seen in Fig. \ref{fig:dac_principle}. In a binary-weighted configuration, the less significant bit (LSB) controls one inverter, while the most significant bit (MSB) controls $2^{N-1}$ inverters in parallel, with $N$ being the number of bits.

The PMOS and NMOS transistors of the inverter will be seen as switches with an associated on-resistance. The output voltage can be calculated as:
\begin{equation}\label{eq:vdac}
    V_{DAC}(m) = \frac{(D_{max}-D_m) g_{op}}{(D_{max}-D_m)g_{op}+D_m g_{on}} \cdot V_{DD}
\end{equation}

where $g_{op}=1/r_{op}$ and $g_{on}=1/r_{on}$ are the on-transconductance values of the PMOS and NMOS transistors, respectively. $D_{max}$ is the maximum number of levels of the DAC ($2^{N}-1$), and $D_m$ is the decimal equivalent of the binary word ($b_{N-1}\ldots b_1b_0$) to be converted from digital to analog.

If both transistors are assumed to have the same on-resistance\footnote{It is common practice to design inverters to have similar dynamic characteristics for high-to-low and low-to-high voltage transitions.}, and this on-resistance do not depend on the voltage across it, the ideal DAC voltage can be obtained as:
\begin{equation}\label{eq:vdac_ideal}
    V_{DACo}(m) = \frac{D_{max}-D_m}{D_{max}} \cdot V_{DD}
\end{equation}

From equation (\ref{eq:vdac_ideal}), and Fig. \ref{fig:dac_principle}, it is clear that the inverter-based DAC behaves like a conventional resistor divider DAC, in the ideal case. Notice that regarding the digital representation with inverters, the DAC voltage is inverted.

\begin{figure}[!t]
	\centering
	\includegraphics[width=1.0\linewidth]{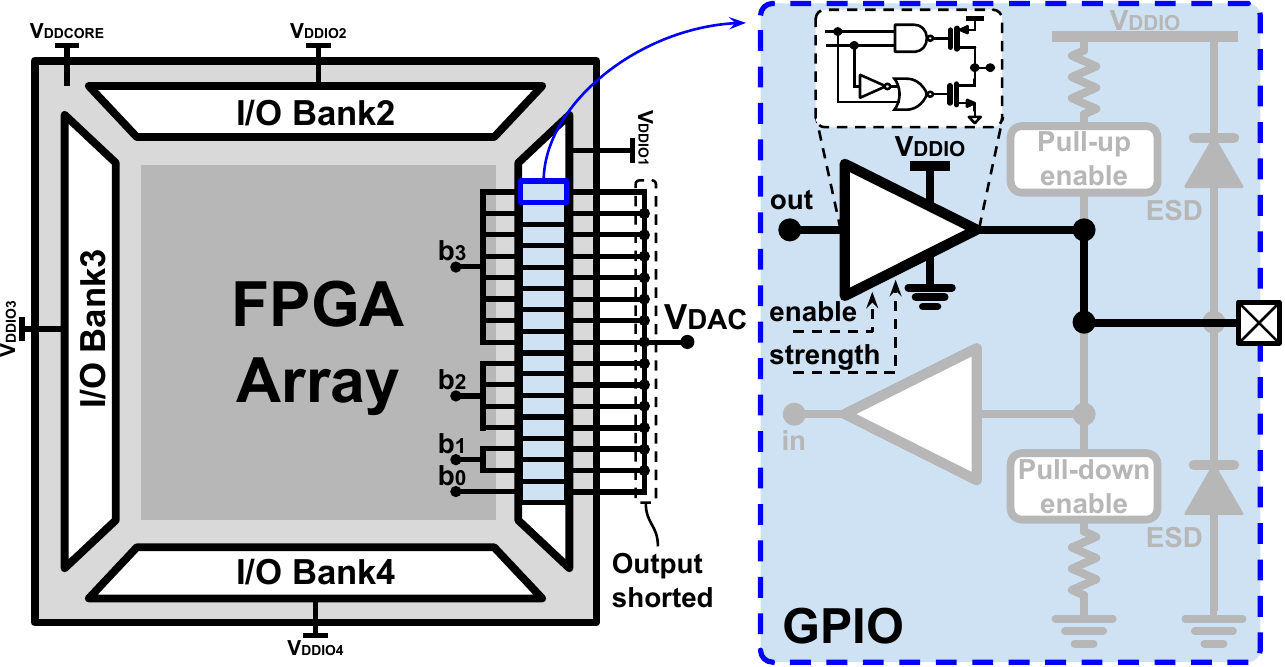}
    \caption{\small FPGA-based DAC with a 4-bit binary-weighted implementation.}	
    \label{fig:fpga_dac}
\end{figure}

Fig. \ref{fig:fpga_dac} shows the proposed FPGA-based DAC without external components. The basic idea is to use the GPIOs of the FPGA in the same fashion as the inverters in the DAC of Fig. \ref{fig:dac_principle}. The proposed scheme shorts the outputs of the GPIOs in order to apply the inverter-based DAC principle. Although we introduce here a new approach to implement an FPGA-based DAC, the idea extends to any device where GPIOs are available, for example, in a microcontroller like the ones used in popular developing boards, such as Arduino, or small single-board computers, like the Raspberry-Pi family.

A general GPIO scheme is shown in Fig. \ref{fig:fpga_dac} as well. For the implementation, the output buffers of the GPIOs are enabled, while the input buffer, as well as the pull-up and pull-down resistors, are disabled.\footnote{One may think that the pull-up and pull-down resistors could be used as an ideal resistor DAC. Unfortunately, their enable pins can not be controlled through internal FPGA's signals, but are hard-configured when synthesizing the circuit in the FPGA. Furthermore, not all FPGAs have both pull-up and pull-down resistors, but just one of them.} Since the output buffer of the GPIO is conformed by two inverters in cascade, the output of the DAC is no longer inverted, and the ideal DAC voltage is no longer given by equation (\ref{eq:vdac_ideal}), but:
\begin{equation}\label{eq:vdac_ideal2}
    V_{DACo}(m) = \frac{D_m}{D_{max}} \cdot V_{DD}
\end{equation}

In contrast with other FPGA-based DACs, this implementation does not need the use of any external component. Furthermore, since the configurable GPIO output buffers' strength dictates the maximum frequency, the proposed scheme might achieve higher frequency operation in comparison to other FPGA-based DACs.

\subsection{Problems and non-idealities}\label{sec:problems}
Even though the FPGA-based DAC implementation has certain advantages, it has some practical issues. The first one being the power consumption. Since the output buffer of a GPIO is designed to handle relative large capacitive loads ($>$50pF) while maintaining appropriate frequencies, the output current of a GPIO is usually significant ($>$10mA) when transitioning from high-to-low, or vice-versa. Taking into account Fig. \ref{fig:dac_principle}, the latter translates to have low on-resistance values. By shorting outputs of the GPIOs, we placed a direct path from the supply voltage to ground through the inverter chain. Hence, the FPGA-based DAC consumes high-static current values, which finds its peak at DAC's output mid-range. The current consumption doubles for each bit of resolution added to the DAC, making it only practical for lower resolution implementations ($<$6-bits). Although some GPIOs also have the capability of adjusting their output strength, still the amount of static-current is considerable.

The other problem is the DAC's linearity. In the ideal inverter case, the on-resistance of the PMOS and NMOS transistors are the same. Although this may be true, the on-resistances are equal only for a small region of the transfer function. To understand how the DAC's transfer function behaves, one has to remember that, in reality, the devices used for the construction of the FPGA-based DAC are not resistors, but transistors, which have a non-linear behavior. From equation (\ref{eq:vdac}), and without the assumption that the on-transconductances $g_{op}$ and $g_{on}$ are equal, an error term can be found:
\begin{equation}\label{eq:vdac2}
V_{DAC}(m) = V_{DACo}(m)\cdot\bigg[1 + \frac{(D_m/D_{max})\cdot \epsilon}{1-(D_m/D_{max})\cdot \epsilon} \bigg]
\end{equation}

where $\epsilon=1-g_{on}/g_{op}$ is the relative error factor of the on-transconductances of the NMOS and PMOS transistors. A Verilog-A model, using a transistor's first-order approximation, was implemented to show the non-linear behavior of the DAC's transfer function, and the effect of the relative error factor.

\begin{figure}[!t]
	\centering
	\includegraphics[width=0.70\linewidth]{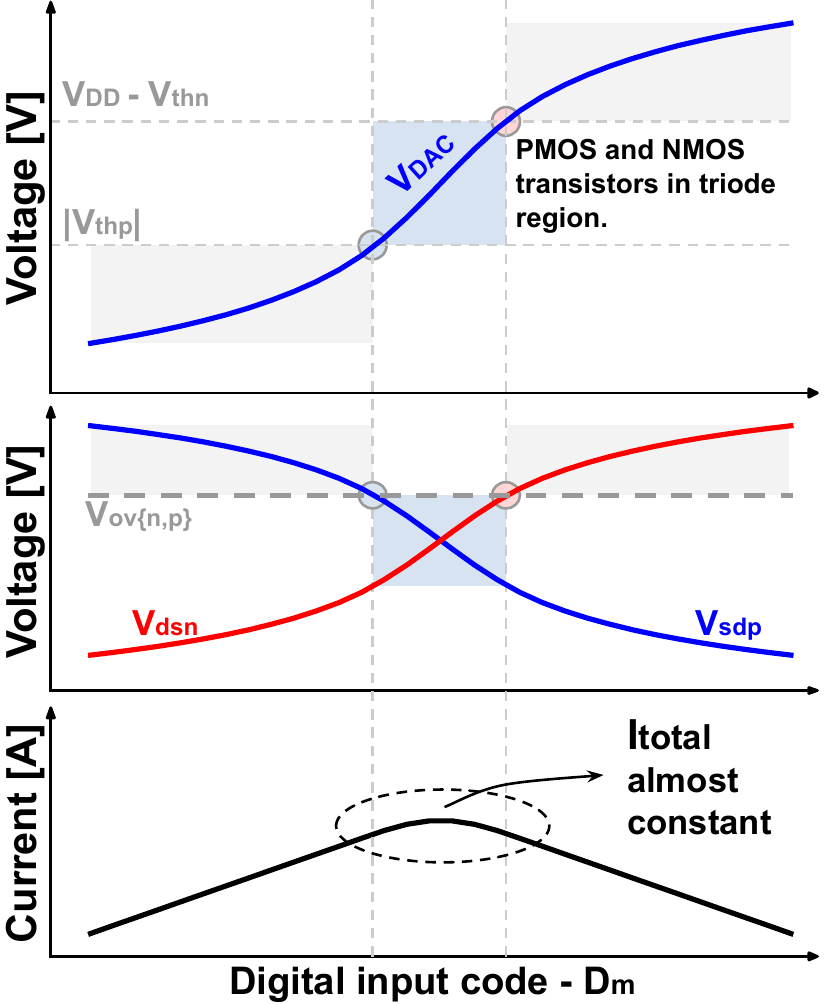}
    \caption{FPGA-based DAC transfer function behavioral illustration (top). Curves obtained from a Verilog-A model simulation. The figure also shows curves of polarization conditions (middle) and current consumption behavior (bottom) for the whole digital input range.}
    \label{fig:dac_uncorr}
\end{figure}

To simplify the analysis, and as already mentioned, the following discussion will be based on the assumption that the PMOS and NMOS transistors have equivalent static and dynamic characteristics. In the Verilog-A model simplest form, the latter translates to two equivalencies: that their threshold voltages are equal in magnitude ($V_{th}=V_{thn} = |V_{thp}|$), and that their transconductance parameters are equal as well ($\mu_pC_{ox}(W/L)_p = \mu_nC_{ox}(W/L)_n$). With this in mind, Fig. \ref{fig:dac_uncorr} depicts the behavioral simulation results for illustration purposes.

When the digital input code is in the low-end of its range, the DAC will have more NMOS active transistors than PMOS active transistors, which will pull-down the output voltage $V_{DAC}$. At this point, since $V_{DAC}$ (NMOS drain-source voltage $V_{dsn}$) is lower than the transistors' overdrive voltage ($V_{ov}$=$V_{ovn}$=$V_{ovp}$, which is constant across the input range and equals to $V_{DD}-V_{thn}$), the NMOS transistors will be in the triode region (see the middle curve of Fig. \ref{fig:dac_uncorr}). Since the source-drain PMOS voltage behavior is complementary to the NMOS one ($V_{sdp}$=$V_{DD}-V_{DAC}$), the PMOS transistors are in saturation ($V_{sdp}$$>$$V_{ov}$). When increasing the number of PMOS active transistors (while proportionally reducing the number of NMOS transistors connected), the $V_{DAC}$ output voltage increases to the point where the PMOS are no longer in saturation. In this region, the NMOS and PMOS transistors are in triode. It is in the triode region where their on-transconductances behave in a similar fashion, hence, it is the most linear region in the transfer function, as it can be appreciated in Fig. \ref{fig:dac_uncorr} (top). Finally, $V_{DAC}$ continues its rise, entering the region where the NMOS transistors are in saturation, while the PMOS transistors are in triode. 

Even though power consumption limits the maximum number of bits that can be used, and that this DAC is not the most linear one, the FPGA-based DAC may be used in many applications. Examples of these applications are: those where linearity is not an inconvenient, but a desired feature \cite{Lu2008}; in feedback loop systems \cite{Custodio2013}; to test ideas where analog circuits are implemented with digital cells, i.e., digital low drop-out regulators implementations \cite{Huang2018}; or in general, when using FPGAs that does not have internal DACs. Furthermore, the DAC's output can be characterized and corrected digitally, or its non-linearity behavior could be taken into account in the systems transfer function. 

In any case, and if the application requires it, a simple method to linearize the DAC's output, while solving the power consumption problem in the process, is presented in the following section. 

\subsection{Linearity correction and current reduction}\label{sec:solutions}
With the qualitative description of the previous section, the non-linear behavior of the DAC's transfer function can be understood. It is important to identify what are the conditions to be in the linear region, in order to apply the necessary corrections. There are two ways of looking at these conditions for linearity. The first one was already discussed in the previous section, and it is the most evident from Fig. \ref{fig:dac_uncorr} (top), as well as a consequence from equation (\ref{eq:vdac2}) and the relative error factor $\epsilon$ of the on-transconductances: if both transistors are in the triode region, they will behave so similar that $\epsilon \approx 0$, hence, the DAC's output would approximate to its ideal value. The second way to look at this is to see the cumulative current behavior in Fig. \ref{fig:dac_uncorr} (bottom). In the linear region, the current of the DAC is almost constant but decays the closer it gets to the digital input range edges. 

\begin{figure}[!t]
	\centering
	\includegraphics[width=1.0\linewidth]{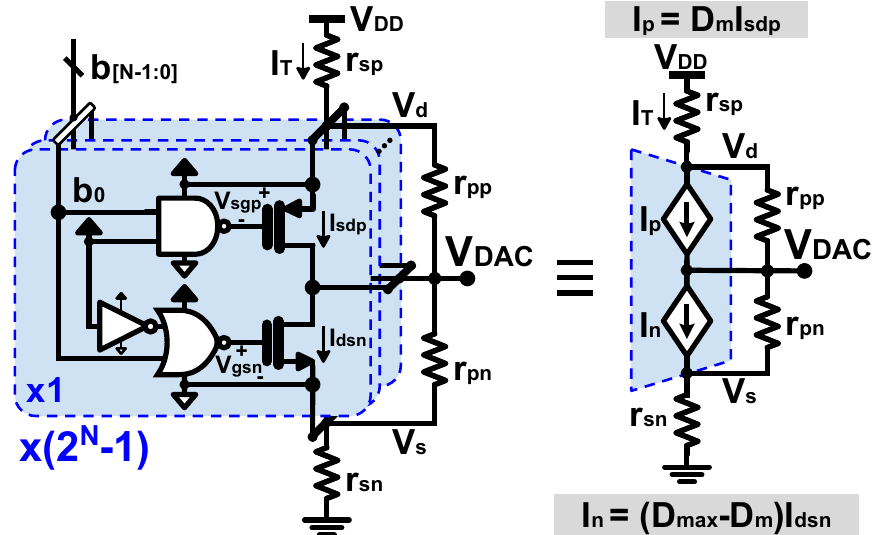}
    \caption{FPGA-based DAC configuration to linearize its output. By placing 2-4 extra resistors, the output can be linearized at the cost of dynamic range loss.}
    \label{fig:dac_correction}
\end{figure}

Then, to improve the DAC's linearity, it is necessary to stretch the triode-triode region condition through the entire digital input range, or/and making the DAC's current constant throughout the whole range. This can be done with the general configuration of Fig. \ref{fig:dac_correction}. Two configurations will be analyzed: resistors in parallel only (two resistors, with $r_{s\{p,n\}}=0$), and the general configuration including the series resistors (four resistors).

\begin{figure}[!t]
	\centering
    \includegraphics[width=1\linewidth]{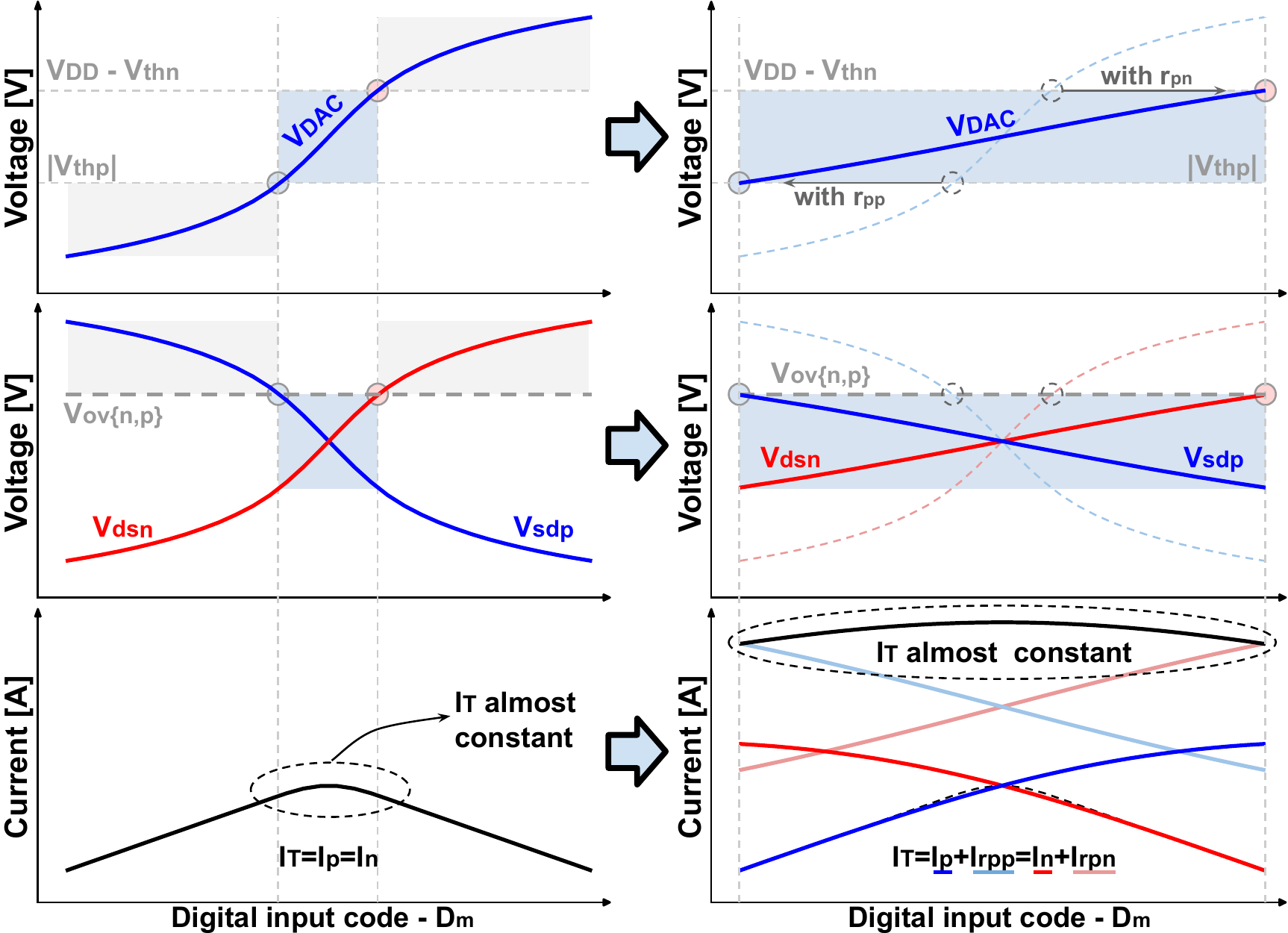}
    \caption{Simulated behavior comparison: the curves in the left show the results for the DAC without any external components, while the curves in the right were obtained by using external resistors in parallel to improve the DAC's linearity.}
    \label{fig:Uncorr_vs_CorrRp}
\end{figure}

The two resistor configuration improves the linearity by including two external resistors, one between $V_{DD}$ and $V_{DAC}$ ($r_{pp}$), and the other between $V_{DAC}$ and ground ($r_{pn}$). The inclusion of the parallel resistors creates additional current paths than those of the main ones (through the GPIOs). To see the parallel resistor effect, let's take into consideration the PMOS transistors with the resistor $r_{pp}$. The $r_{pp}$ path (in a certain way) complements the main path, such that the sum of their currents is almost constant (which depends on the $r_{pp}$ resistance value), as it can be appreciated in Fig. \ref{fig:Uncorr_vs_CorrRp} (bottom). In this way, when there are fewer PMOS active transistors, the remaining current flows through the additional path created by the additional $r_{pp}$. The same happens between the NMOS transistors and the resistor $r_{pn}$.

From another point of view, the triode-triode region stretches out by connecting the parallel resistors. By adding the two parallel resistors to the expression in (\ref{eq:vdac2}), one can obtain:
\begin{equation}
V_{DAC}(m) = \frac{D_m + g_{pp}/g_{op}}{D_{max}-D_{m}\cdot\epsilon + (g_{pp}+g_{nn})/g_{op}} \cdot V_{DD},
\end{equation}
where $g_{p\{p,n\}}=1/r_{p\{p,n\}}$. Assuming that the parallel resistors have the same value ($r_{pp}=r_{pn}$), and that in the linear region the on-transconductances of the GPIOs' NMOS and PMOS transistors are approximately the same ($g_{op}\approx g_{on}$, or $\epsilon\approx 0$), then:
\begin{equation}
V_{DAC}(m) \approx \frac{D_m + \alpha_g}{D_{max} + 2\alpha_g} \cdot V_{DD},
\end{equation}
where $\alpha_g=g_{p\{p,n\}}/g_{o\{p,n\}}$. With the previous equation, it is clear that by selecting the proper $\alpha_g$ value, the linear region can be stretched out to the entire digital input range, or in other words, it can be stretched out from $D_m=0$ by setting $V_{DAC}=|V_{thp}|$, to $D_m=D_{max}$ by setting $V_{DAC}=V_{DD}-V_{thn}$. With the latter, it is easy to estimate the relation between the parallel resistance value and the transistors' on-resistance as:
\begin{equation}\label{eq:Rparallel1}
\alpha_g = \frac{r_{o\{p,n\}}}{r_{p\{p,n\}}} \approx \frac{D_{max}\cdot V_{th}}{V_{DD}-2V_{th}}
\end{equation}

All necessary variables in the previous equation can be found experimentally when obtaining the DAC's transfer function. For example, the $V_{DD}-2V_{th}$ is the linear's region dynamic range (see the top plot in Fig. \ref{fig:Uncorr_vs_CorrRp}), from which the transistors' threshold voltage $V_{th}$ can be derived (knowing the supply voltage $V_{DD}$), while the on-resistance value can be derived by measuring the current ($I_{o\{p,n\}}$) through one GPIO at the input's mid-range.

The two resistor configuration for linearity correction is easy enough to implement since only two external resistors are needed. Not complicated schemes are applied to obtain a more linear transfer function. The relevant issue is its implementation practicality. Fig. \ref{fig:Uncorr_vs_CorrRp} summarizes that by adding resistors in parallel, the total current consumption of the system is larger (almost as twice than without the correction). This is even more detrimental when increasing the number of bits of the DAC's implementation. Besides, the dynamic range of the DAC decreases, which might be no critical depending on the application. It is important to emphasize that the number of bits holds throughout a lower dynamic range, which translates to a higher resolution DAC.

\begin{figure}[!t]
	\centering
    \includegraphics[width=1\linewidth]{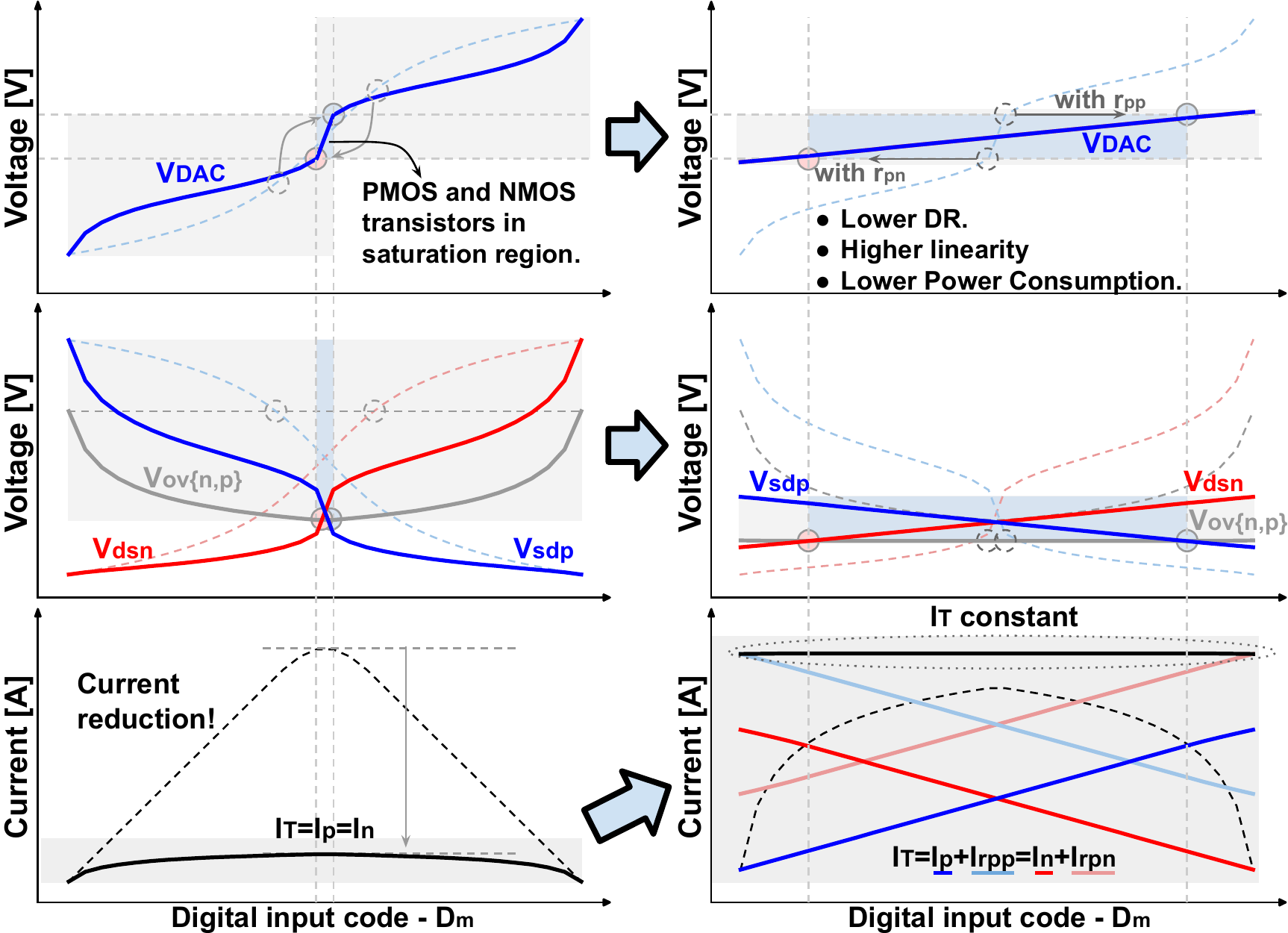}
    \caption{Simulated model with linearity correction. The curves in the left show the effect of the series resistors, while the curves in the right were obtained by adding parallel resistors to complete the linearity correction configuration.}
    \label{fig:Uncorr_vs_CorrAll}
\end{figure}

Up to now, the discussion regarding the linearity has focused on the condition that the NMOS and PMOS transistors should be in the triode-triode region. In reality, the conclusion from equation (\ref{eq:vdac2}) is not that both have to be in the triode region, but both have to behave, if not the same, as similar as possible. In fact, better linear performance could be obtained if the transistors are in saturation since they will be almost independent of the drain-source voltage\footnote{Neglecting channel-length modulation.}. With the configurations that have been analyzed until now, a region where both group of transistors could be in saturation at the same time is not possible, since their gate and source voltages are always the same ($V_{gsn}=V_{sgp}=V_{DD}$). This is no longer the case if the general four resistor configuration of Fig. \ref{fig:dac_correction} is considered.

First, the effect of the series resistors will be analyzed. By adding these resistors, the $V_{d}$ and $V_{s}$ voltages (Fig. \ref{fig:dac_correction}) can be modified, which changes the gate-source absolute voltages ($V_{gsn}=V_{sgp}=V_d-V_s$). Hence, the overdrive voltage ($V_{ov}=V_{gsn}-V_{thn}=V_{sgp}-|V_{thp}|$) is changed as shown in Fig. \ref{fig:Uncorr_vs_CorrAll} (left-center), reducing it at mid-range. By increasing the series resistance value, it is possible to find a region where both transistor groups are in saturation ($V_{dsn}$$\geq$$V_{ov}$ and $V_{sdp}$$\geq$$V_{ov}$), as it is illustrated in Fig. \ref{fig:Uncorr_vs_CorrAll} (left). Furthermore, and as expected, the addition of the series resistors had the advantage of decreasing the current consumption as shown in Fig. \ref{fig:Uncorr_vs_CorrAll} (left-bottom).

Finally, by adding the parallel resistors, the circuit will experience the same effect as with the two resistors configuration, but the region to be stretched out is the saturation-saturation region, as shown in Fig. \ref{fig:Uncorr_vs_CorrAll} (right). Although the results are similar to the two resistor configuration, three differences can be noticed. The first one is that the four resistor configuration seems to have a linearity improvement, which can be explained by both group of transistors being in the saturation region rather than in triode, as commented before. The second one is that the dynamic range is somehow lower for the four resistor configuration. The dynamic range is heavily dependent on the transistors' threshold voltage: for the two resistor configuration, the lower the $V_{th}$ the better, while for the four resistor configuration is all the way around. The final difference is the total current of the system. Values as 10X lower than the standalone (no resistors) or two resistor configurations can be obtained. In turn, the series-parallel configuration is slower than the other two, as expected.

Now, to find the resistors' values, the circuit in Fig. \ref{fig:dac_correction} (right) will be analyzed such that the transistors are in strong inversion\footnote{Some simulations have shown that it is possible to apply the same concept in weak inversion, but this will not be discussed in this work.} ($V_{gsn}$$\geq$$V_{thn}$ and $V_{sgp}$$\geq$$|V_{thp}|$) and saturation condition ($V_{dsn}$$\geq$$V_{ov}$ and $V_{sdp}$$\geq$$V_{ov}$). With the latter in mind, the following inequalities can be found:
\begin{equation}\label{eq:condition}
\begin{gathered}
    V_{DD}-I_{T}\cdot (r_{sp}+r_{sn}) \geq |V_{th\{n,p\}}|\\
    V_{DD}-I_{T}\cdot r_{sp}-V_{thn}\leq V_{DAC} \leq I_{T}\cdot r_{sn} + |V_{thp}|
\end{gathered}
 \end{equation}

Two things can be derived from the previous inequalities: the maximum and minimum values of the DAC's output ($V_{DAC_{max}}$ and $V_{DAC_{min}}$), and the limits of the total current $I_T$ (supposing $V_{th}=V_{thn}=|V_{thp}|$):
\begin{equation}\label{eq:Itotal}
    \frac{V_{DD}-2V_{th}}{r_{sp}+r_{sn}} \leq I_T \leq \frac{V_{DD}-V_{th}}{r_{sp}+r_{sn}}
\end{equation}

The latter inequality can be used to find the series resistors' values if a total current specification is set and if both resistors are assumed to have the same value for symmetry. Finally, to find the parallel resistors' values, the circuit in Fig. \ref{fig:dac_correction} (right) is solved to obtain the following pair of equations:
\begin{equation}\label{eq:dac_circuit1}
    V_{DAC} = V_{DD} -I_{T}\cdot (r_{sp}+r_{sn}) + I_p \cdot r_{pp}
\end{equation}
\begin{equation}\label{eq:dac_circuit2}
    V_{DAC} = I_{T}\cdot (r_{sp}+r_{sn}) - I_n \cdot r_{pn}
\end{equation}

where $I_p=D_m\cdot I_{sdp}$ and $I_n=(D_{max}-D_m)\cdot I_{dsn}$, as stated in Fig. \ref{fig:dac_correction}. Replacing $V_{DAC_{min}}$ and $D_m=0$ in equation (\ref{eq:dac_circuit1}), and $V_{DAC_{max}}$ and $D_m=D_{max}$ in equation (\ref{eq:dac_circuit2}), the parallel resistors' values can be found as:
\begin{equation}\label{eq:Rparallel2}
    r_{p\{p,n\}}=\frac{|V_{th\{n,p\}}|}{I_T}
\end{equation}

All the variables needed to find the solutions to equation (\ref{eq:condition}) and equation (\ref{eq:Rparallel2}) can be obtained experimentally. 

In this section, it has been shown to be theoretically possible to obtain an FPGA-based DAC without any external components. Far from perfect, current and linearity are the two major drawbacks of the standalone implementation. Still, the DAC could be used for several applications. In any case, two different configurations were proposed to solve such problems by only adding a few external resistors, obtaining interesting results which will be verified in the following (experimental) section. 

\section{Experimental results}\label{sec:results}

\begin{figure}[!t]
	\centering
    \includegraphics[width=0.8\linewidth]{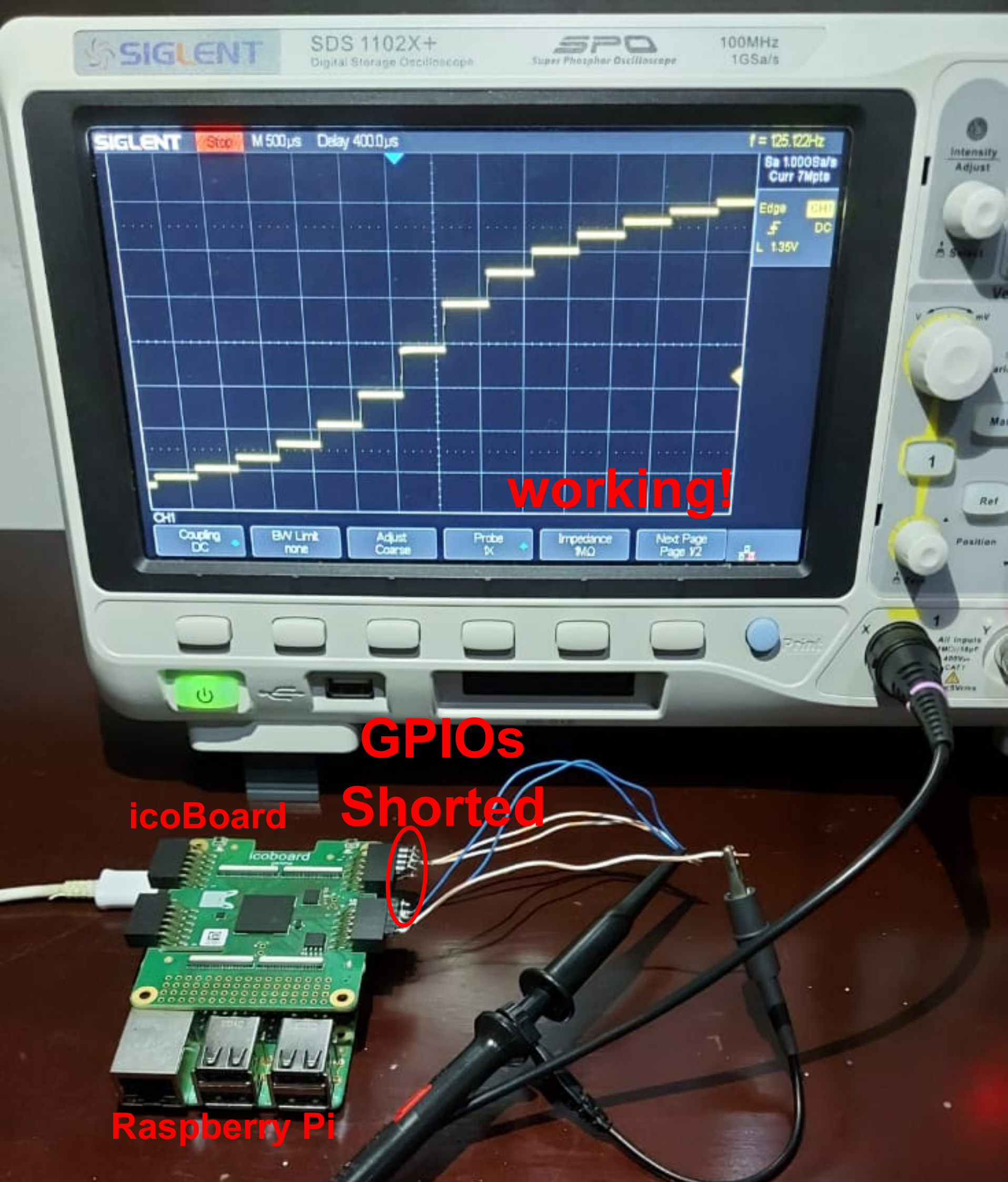}
    \caption{Measurement setup used to validate the FPGA-based concept. An icoBoard v1.1 was programmed and powered through a Raspberry Pi 3 model B.}
    \label{fig:testbench}
\end{figure}

This section presents the measurement results performed to validate the theoretical foundation described in the previous section. For this, a testbench using the icoBoard v1.1, which contains a Lattice FPGA, 100MHz max clock, up to 8 MBit of SRAM, up-to 206 GPIOs, and it is programmable in Verilog by a complete open-source FPGA toolchain (consisting of Yosys and ArachnePnR and icetools) \cite{icoboard}. The icoBoard is programmed and powered through a Raspberry Pi 3 model B\footnote{Although the concept could have been implemented directly over the Raspberry Pi board, it is simpler to test the concept in an FPGA due to its versatility.}. The complete setup can be seen in Fig. \ref{fig:testbench}.

\begin{figure}[!t]
	\subfigure[DAC's output (modified oscilloscope's image).]
	{
    	\centering
        \includegraphics[width=0.95\linewidth]{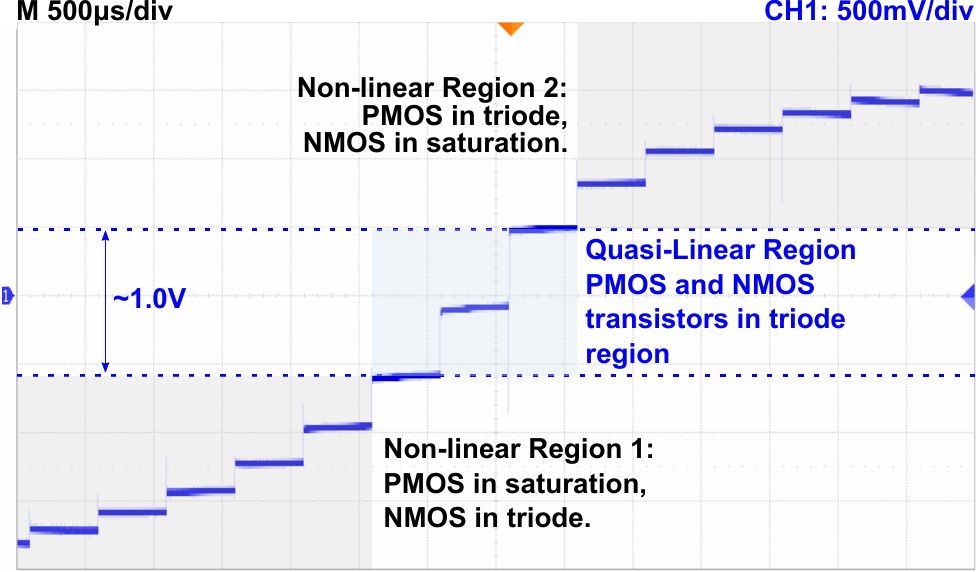}
        \label{fig:DAC4b_transfer}
    }
    \vskip -0.5\baselineskip
    \subfigure[DAC's total current consumption.]
	{
    	\centering
        \includegraphics[width=0.95\linewidth]{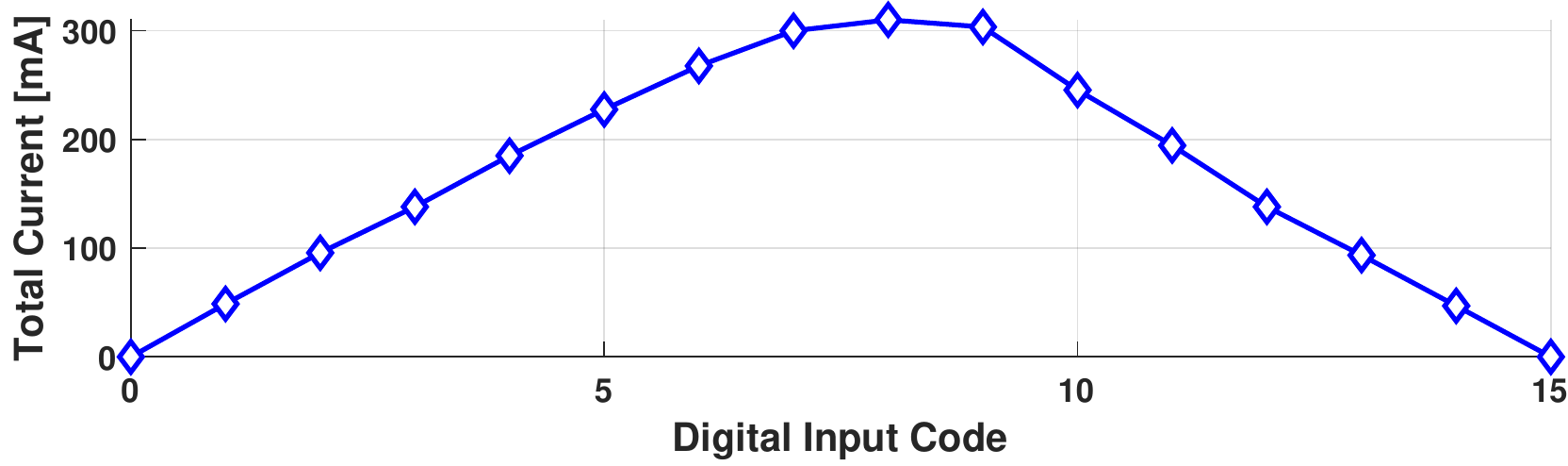}
        \label{fig:DAC4b_Itotal}
    }\vspace{-1\baselineskip}
\caption{Measurement results for a 4-bit FPGA-based DAC implementation.}
\label{fig:DAC4b_results}
\end{figure}

The following measurements were performed by implementing a 4-bit DAC standalone configuration. Fifteen GPIO outputs were shorted-out as in Fig. \ref{fig:fpga_dac}. To see the DAC's output in an oscilloscope, a low-frequency periodical stair-case was programmed in the FPGA, and the results are shown in Fig. \ref{fig:DAC4b_transfer}. The transfer function goes from 0V to 3.3V ($V_{DD}$), and the sampling period was set to $\sim$500$\mu$s. As explained in section \ref{sec:problems}, the FPGA-based DAC's output has two non-linear regions and a (quasi) linear region at mid-range, which are shown in the figure. The linear region has a dynamic range of $\sim$1V, which in reality could be greater, but since the number of steps in this region is limited to two digital codes, the dynamic range measurement is not the most accurate. 

The DAC's total current was measured as well and presented in Fig. \ref{fig:DAC4b_Itotal}. The current consumption peak occurs at mid-range, and it is approximately 300mA, which would give $\sim$40mA per GPIO. A 5-bit DAC configuration was still possible, which would consume almost as twice the current ($\sim$600mA), but it would have been a problem when trying to implement the two configurations for linearity correction, which would have consumed around four times the current of a standalone 4-bit implementation ($\sim$1.2A). Also, it should be noted that the maximum DC current that a single GPIO can drive is rated at 24mA \cite{lattice_fpga}. Hence, the standalone configuration is limited at low frequencies by this maximum.

We extract the variables needed for the equations shown in section \ref{sec:solutions} from Fig. \ref{fig:DAC4b_results}. For instance, the transistors' threshold voltage can be derived from the linear region's dynamic range, which would be $V_{DD}-2V_{th}\approx 1$V. Knowing that the supply voltage is 3.3V, the threshold voltage would be $\sim$1.15V (as stated before, the linear region's dynamic range could be higher, hence $V_{th}$$\leq$1.15V). Another important variable is the transistors' on-resistance value. It is possible to obtain this value from Fig. \ref{fig:DAC4b_Itotal}, within the linear region. The on-resistance value is calculated to be $r_{o\{p,n\}}\approx40\Omega$.

\begin{figure}[!t]
	\centering
    \includegraphics[width=0.95\linewidth]{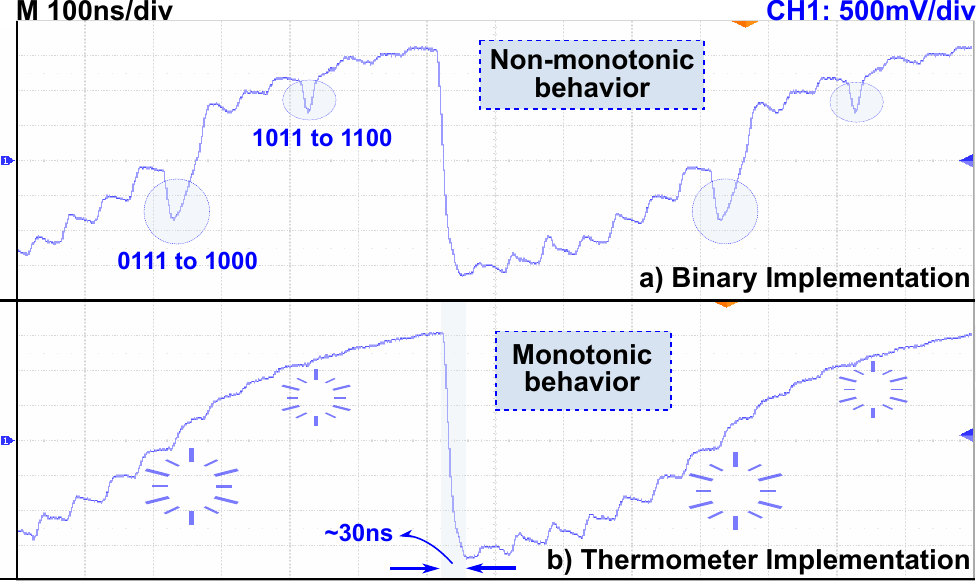}
    \caption{Measurements for a 4-bit DAC at higher frequencies: (a) binary implementation, (b) thermometer implementation.}
    \label{fig:bin_thermo}
\end{figure}

On the other hand, the maximum frequency is limited by the GPIO's dynamic performance, e.g., rising and falling times. Fig. \ref{fig:bin_thermo}(a) shows the result for a sampling rate of 20MS/s of a binary DAC implementation. As it can be seen, this result describes a non-monotonic behavior at abrupt bit changes. To avoid the non-monotonicity, a thermometer implementation can be used, as shown in Fig. \ref{fig:bin_thermo}(b). Higher sampling rates can be obtained according to the measured high-to-low transition time ($\sim$30ns). The transition time of the standalone DAC is the same as the measured transition time of a single GPIO under normal FPGA operation. The latter may lead to the conclusion that the dynamics characteristics of the DAC is dominated by that of the GPIO, but the dynamic measurements for this setup are heavily dependent on the oscilloscope's probe load, which is around 85pF-120pF. It is expected to have faster responses in scenarios where the GPIOs are not as heavily loaded. In any case, the possibility to go as fast as the GPIO's dynamic capabilities is definitely one of the advantages of the proposed FPGA-based DAC compared to other multi-bit implementations.

\begin{figure}[!t]
    \subfigure[DAC's output for the two resistor configuration.]
	{
    	\centering
        \includegraphics[width=1\linewidth]{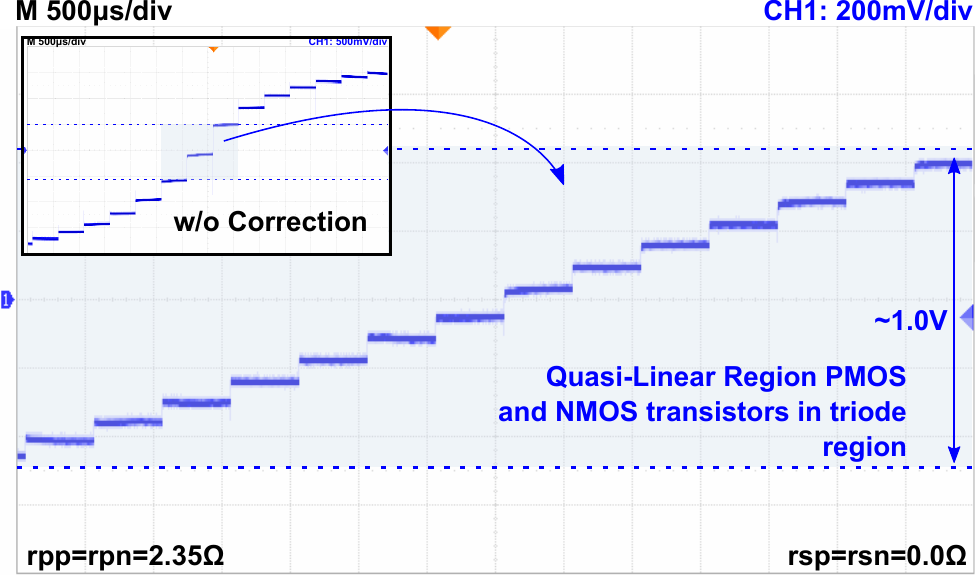}
        \label{fig:DAC4b_Rp_transfer}
    }\vskip -0.5\baselineskip
    \subfigure[Measured DNL (top) and INL (bottom) without external components (gray lines) and with parallel resistors (blue lines).]
	{
    	\centering
        \includegraphics[width=1\linewidth]{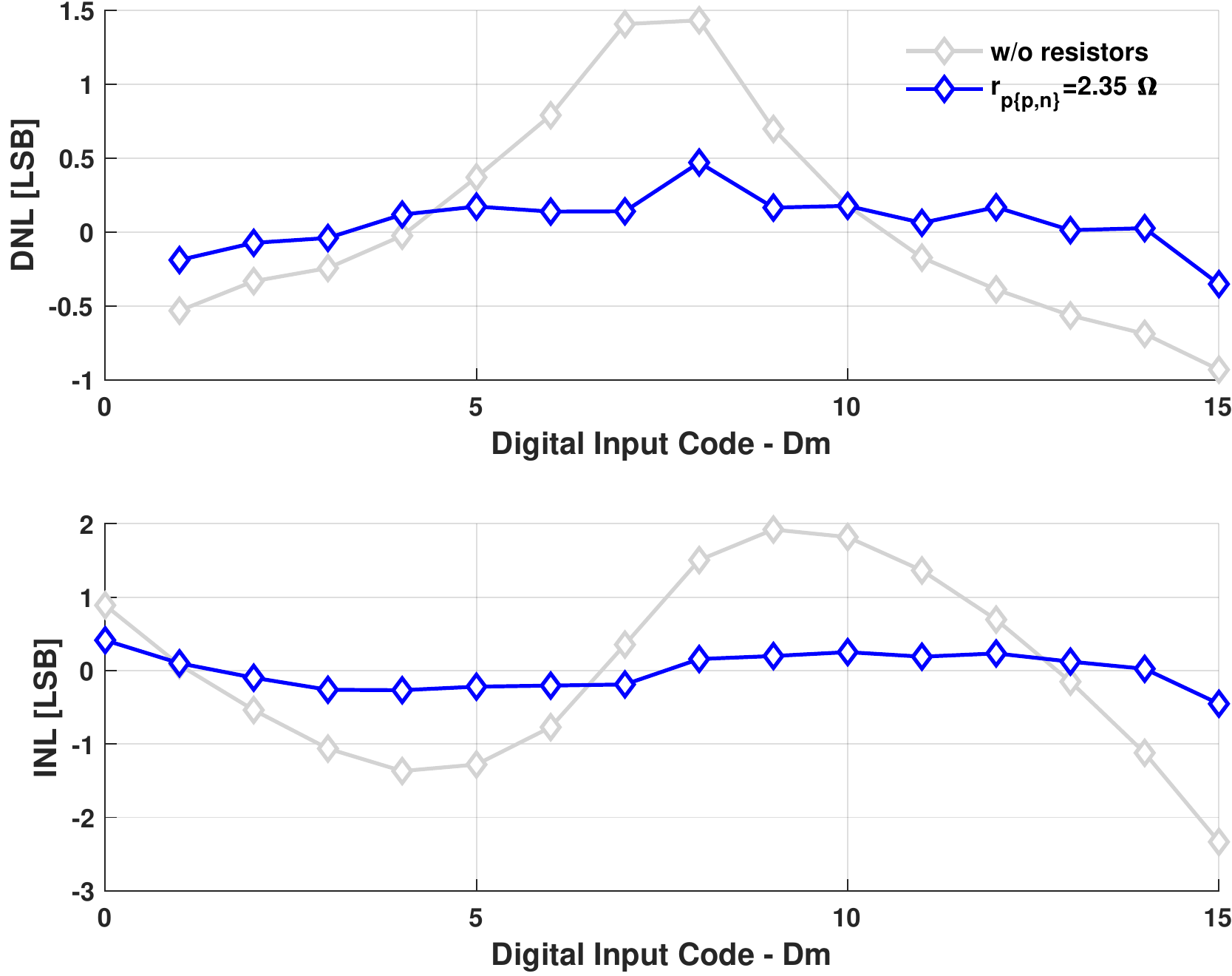}
        \label{fig:DAC4b_UnCorr_vs_CorrRp}
    }\vspace{-1\baselineskip}
\caption{Measurements for a 4-bit DAC using parallel resistor correction only. The dynamic range is reduced, but the DAC is linear within the whole range.}
\label{fig:DAC4b_Rp}
\end{figure}

In section \ref{sec:solutions},  configurations were proposed to correct the non-linearity behavior. For this, two to four external resistors are needed. Fig. \ref{fig:DAC4b_Rp} shows the results of 4-bit DAC when using the two resistor configuration $r_{pp}$ and $r_{pn}$ as shown in Fig. \ref{fig:dac_correction} (right). From equation (\ref{eq:Rparallel1}), it is possible to calculate the parallel resistance value that is needed with the $V_{th}$ and $r_{o\{p,n\}}$ values calculated before. The latter gives a resistance of $\sim$2.3$\Omega$. In the implementation, 2.35$\Omega$ resistors were used.\footnote{The 2.35$\Omega$ resistor was obtained using two commercial 4.7$\Omega$ resistor.} As it can be seen, the linear region, where the PMOS and NMOS transistors are triode region, is completely stretched out to the entire input range. The linearity improvement is notorious, but to quantify it, Fig. \ref{fig:DAC4b_UnCorr_vs_CorrRp} shows the differential non-linearity (DNL) and integral non-linearity (INL) comparison between the standalone and two resistors configurations. While the standalone DAC has a DNL$\leq$1.5LSB and an INL$\leq$2LSB, the two resistors configuration has a DNL$\leq$0.5LSB and an INL$\leq$0.5LSB.

Knowing that the parallel resistance value is 2.35$\Omega$, at mid-range the current through the resistor is $\sim$700mA, while the current through the transistors at mid-range is $\sim$300mA, giving the round total of $\sim$1.0A, which will be almost constant in the entire input range, as described in section \ref{sec:solutions}. This is more than three times the current in the standalone configuration, which is a problem that could be even worse for higher resolution implementations.

\begin{figure}[!t]
	\subfigure[DAC's output for the series and parallel resistors configuration.]
	{
    	\centering
        \includegraphics[width=1\linewidth]{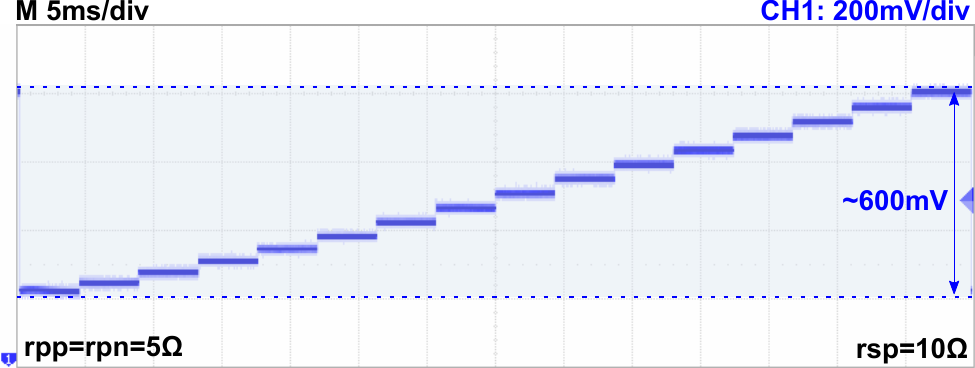}
        \label{fig:DAC4b_RpRs_transfer}
    }\vskip -0.5\baselineskip
    \subfigure[Total current across the input range.]
	{
    	\centering
        \includegraphics[width=1\linewidth]{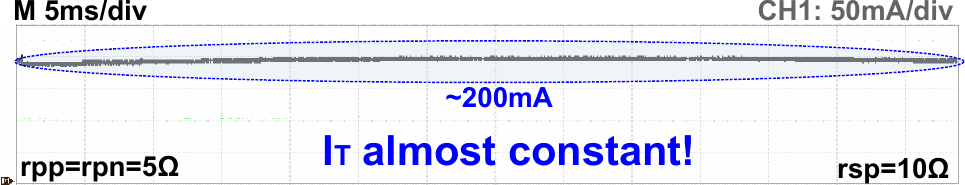}
        \label{fig:DAC4b_RpRs_Itotal}
    }\vskip -0.5\baselineskip
    \subfigure[DNL (top) and INL (bottom) using a 10$\Omega$ series resistor and 5$\Omega$ to 10$\Omega$ parallel resistors.]
	{
    	\centering
        \includegraphics[width=1\linewidth]{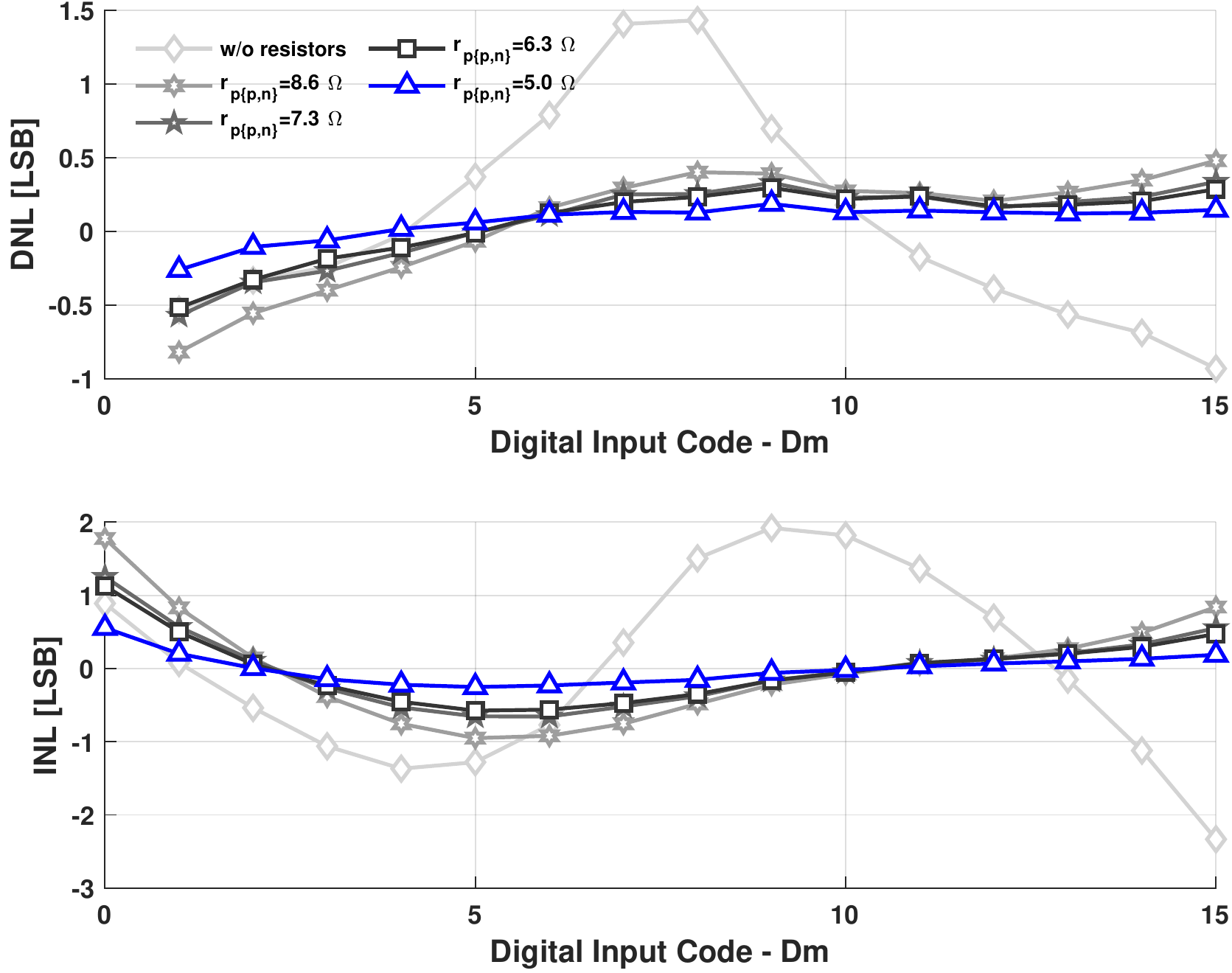}
        \label{fig:DAC4b_RpRs_lin}
    }\vspace{-1\baselineskip}
\caption{Measurements for a 4-bit DAC using series and parallel resistors correction. The DAC is linear within the whole range while the total current is reduced x5 compared to the two resistor configuration.}
\label{fig:DAC4b_RpRs}
\end{figure}

To solve the current consumption problem, a more general four resistor configuration was proposed. For a practical implementation, several things must be considered. For example, it is common that the GPIO banks do not have independent grounds, but the entire FPGA shares the same ground. For the particular setup case, the ground plane of the icoBoard does not allow the inclusion of the $r_{sn}$ of Fig. \ref{fig:dac_correction} (right), hence $r_{sn}=0\Omega$. On the other hand, although each GPIO bank has its independent supply voltage ($V_{DDIOs}$ in Fig. \ref{fig:fpga_dac}), the icoBoard has a single supply plane. To include $r_{sp}$, a cut was done on the PCB immediately after the board's regulator. To allow different series resistor's values, a potentiometer was placed as the $r_{sp}$. 

\begin{figure}[!t]
	\centering
    \includegraphics[width=1.00\linewidth]{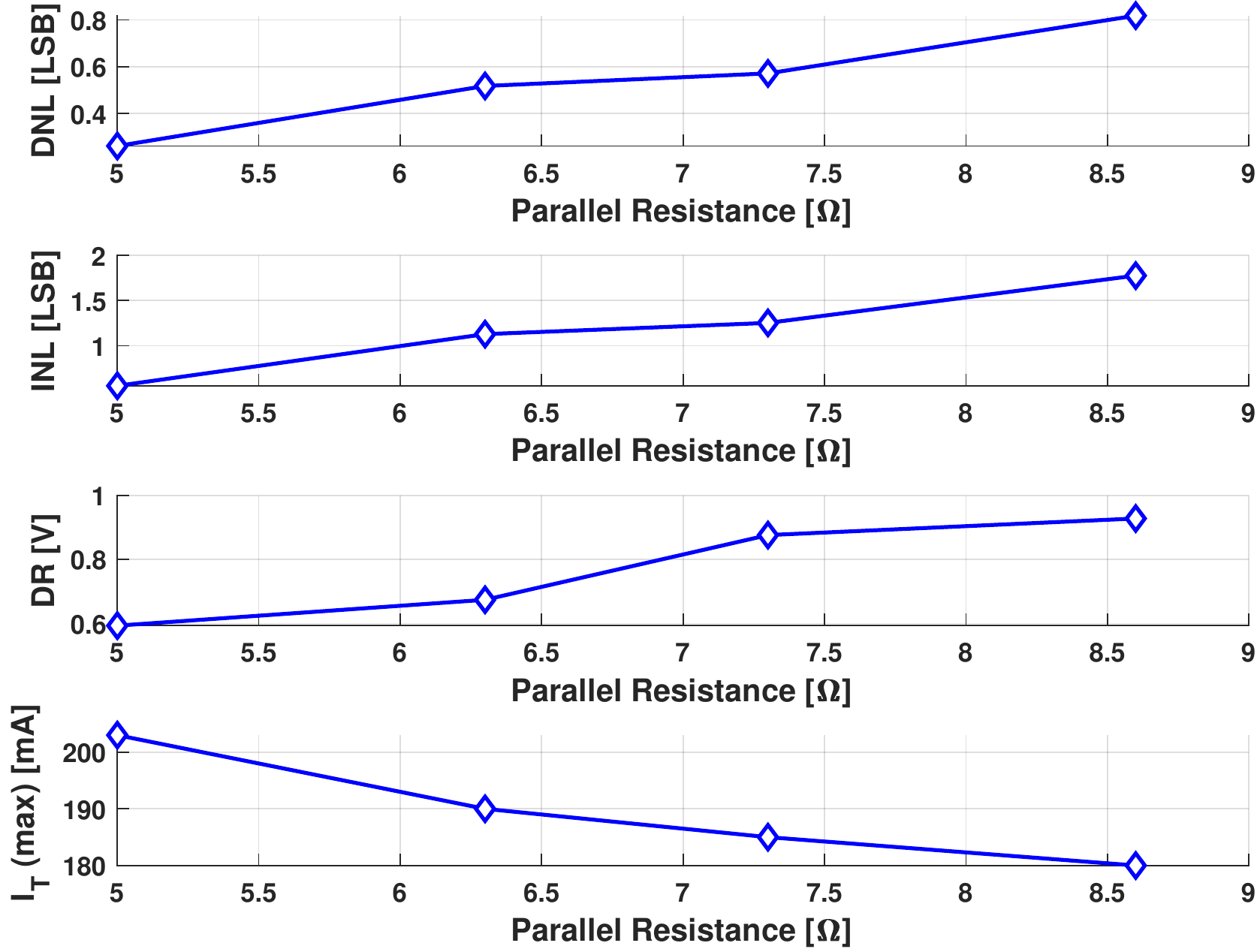}
    \caption{Effect of parallel resistor value on the measured DNL, INL, dynamic range (DR) and total current ($I_T$) for a 4-bit DAC implementation with a 10$\Omega$ series resistor.}
    \label{fig:DAC4b_DR_Itotal}
\end{figure}

Fig. \ref{fig:DAC4b_RpRs_transfer} and Fig. \ref{fig:DAC4b_RpRs_Itotal} show the DAC's output and current consumption, respectively. The 4-bit DAC uses a 10$\Omega$ series resistor and 5$\Omega$ parallel resistors for linearity correction, allowing 5X lower current than the parallel resistor configuration. Theoretically, lower currents can be achieved by using higher resistor values, but since all FPGA's supply pins share the same voltage plane, when putting higher values the FPGA's total current was very limited, causing the FPGA to reset. The impossibility to use higher resistors also limited the dynamic range, and only $\sim$600mV was achieved. Regarding a final application, a careful PCB design should keep the IO supply isolated from the FPGA core, as the chipmaker usually recommends it. With an IO supply isolated, it might be possible to withhold large voltage drops and, therefore, larger supply series resistors. 

Fig. \ref{fig:DAC4b_RpRs_lin} shows the linearity measurement results comparing the standalone and the series-parallel resistors configuration. As expected, the linearity is improved with respect to the parallel resistors configuration, obtaining a DNL$\leq$0.25LSB and an INL$\leq$0.5LSB. Since there is some margin in regards to the linearity, it was decided to test several parallel resistors to improve the dynamic range as showed in Fig. \ref{fig:DAC4b_RpRs_lin}. For a better understanding, the measured DNL, INL, dynamic range, and maximum current are plotted against different parallel resistors values in Fig. \ref{fig:DAC4b_DR_Itotal}. Although there is an improvement in the dynamic range, as well as a reduction in the current, the linearity is worsened considerably. In any case, the final result is more linear than the standalone configuration, hence depending on the application, placing larger resistors may be beneficial.    

\begin{figure}[!t]
	\subfigure[DAC's output.]
	{
    	\centering
        \includegraphics[width=0.95\linewidth]{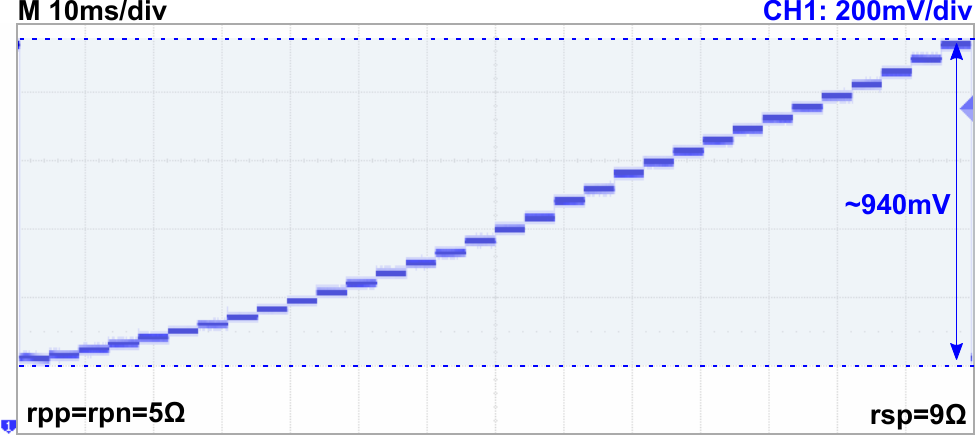}
        \label{fig:DAC5b_RpRs_transfer}
    }\vskip -0.5\baselineskip
    \subfigure[Total current across the input range.]
	{
    	\centering
        \includegraphics[width=0.95\linewidth]{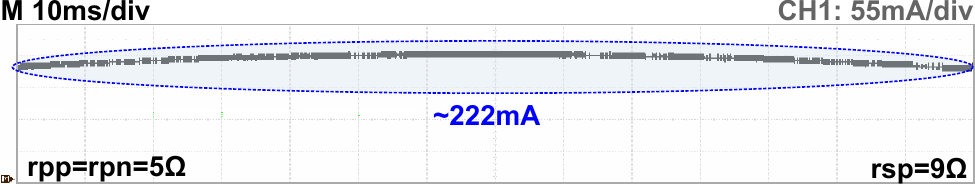}
        \label{fig:DAC5b_RpRs_Itotal}
    }\vspace{-1\baselineskip}
\caption{Measurements for a 5-bit DAC implementation using a series resistor of 9$\Omega$ and parallel resistors of 5$\Omega$ for linearity correction.}
\label{fig:DAC5b_RpRs}
\end{figure}

Finally, it should be noted that the proposed DAC is not limited to a 4-bits configuration. Theoretically, it is only limited by the number of GPIOs, but practically it is limited by the current, as well as other factors, like how the FPGA's PCB has been laid out. To show higher resolution capabilities, a 5-bit implementation using the series-parallel resistors configuration was measured. The DAC's output as well as its current are presented in Fig. \ref{fig:DAC5b_RpRs}. In a standalone configuration, the DAC's maximum current was expected to be twice the current through the 4-bits standalone implementation, but since the series-parallel resistors configuration is used, the current is as low as $\sim$222mA. On the other hand, although the DAC's linearity seems better than a standalone configuration, the best series-parallel resistor setup could not be tested due to higher currents involved which were putting on reset the FPGA.

\section{Summary and Remarks}

In this work, we proposed an all-digital FPGA-based DAC. A general qualitative comparison between different FPGA-based DACs is summarized in Table \ref{table:qual_comp}. The comparison is made with general characteristics of single-bit and other multi-bit DAC implementations. As mentioned throughout the text, the main contribution from the proposed DAC in the standalone configuration is that it does not need external components. As well, higher frequencies can be achieved than other implementations since the standalone proposed DAC depends mainly on the GPIO's dynamic characteristics, while others are limited by the external components used, not to mention the latency added by the required implementation loops (e.g., $\Delta\Sigma$-DAC). 

On the other hand, the standalone DAC implementation does not do very well in other metrics. In particular, the number of GPIOs required for the proposed implementation is the highest of all, which becomes its biggest disadvantage. It is also the worst in resolution, linearity, and power consumption, as it was shown in section \ref{sec:results}. Although these specifications are severely impacted by the external components used in other DAC implementations. For example, the R-2R DAC linearity will be as good as the accuracy of the resistors used for the implementation, which translates to more costs in the DAC implementation. On the other hand, the resolution, linearity, and power consumption of a PWM or $\Delta\Sigma$-DAC depends on the filter used externally. For higher-resolutions implementations, higher-order active filter architectures may be required, which increases the implementation costs, power consumption, and complexity. 

\begin{table}[!t]
\centering
\caption{Qualitative comparison between FPGA-based DACs.}
\label{table:qual_comp}
\begin{threeparttable}
\resizebox{\linewidth}{!}{%
\begin{tabular}{l|cccc}
    &   \textbf{Single-bit}              &   \textbf{Other Multi-bit}               &   \multicolumn{2}{c}{\textbf{Proposed DAC}}\\
    &   \textbf{(PWM, $\Delta\Sigma$)}   &   \textbf{(e.g. R-2R, etc.)}    &   \textbf{(standalone)}   &   \textbf{(with correction)}\\
\hline
\hline
\textbf{External}   &   \multirow{2}{*}{YES (-)}        &   \multirow{2}{*}{YES (- -)}  & \celdaBlue    &   \multirow{2}{*}{YES (+)}\\
\textbf{Components} &                                   &                               & \celdaBlue\multirow{-2}{*}{\textbf{NO (++)}}  &\\
\textbf{Frequency}  &   LOW (- -)                       &   HIGH (+)         & \celdaBlue \textbf{HIGH (++)}    &   LOW (-)\\
\textbf{Resolution} &   \celdaBlue \textbf{HIGH (++)}   &   MEDIUM (+)       & LOW (- -)    &   LOW (-)\\
\textbf{Linearity}  &   \celdaBlue \textbf{HIGH (++)}   &   HIGH (+)         & LOW (- -)    &   MEDIUM (-)\\
\textbf{Power}      &   \celdaBlue \textbf{LOW (++)}    &   MEDIUM (-)       & HIGH (- -)   &   LOW (+)\\
\textbf{Nr. GPIOs}  &   \celdaBlue \textbf{LOW (++)}    &   MEDIUM (-)       & HIGH (- -)   &   HIGH (- -)\\
\hline
\hline
\end{tabular}
}
\end{threeparttable}
\end{table}

To improve the DAC's performance, an entire section was dedicated to discussing the problems and non-idealities of the proposed implementation. From the conclusions obtained, simple configurations were proposed to correct those problems. By including only two to four external resistors, the power consumption was reduced (5X lower), and the linearity was improved considerably (DNL$\leq$0.25LSB and INL$\leq$0.5LSB), at the expense of lower dynamic range. Although the dynamic range is reduced, the effective number of bits is maintained, which in turn improves the DAC's resolution. Reducing the power consumption also allows the possibility to use more number of bits than those shown in this work (4-bits and 5-bits implementations were demonstrated), limited only by the number of GPIOs available in the FPGA. 

To conclude, the fact that the proposed FPGA-based DAC does not need external components enables those FPGAs without internal DACs to be used in analog or mixed-signal systems. But if the application requires it, with the inclusion of
few external resistors, better performance can be achieved in terms of power consumption, linearity, and resolution.

\bibliographystyle{IEEEtran}
\bibliography{References}

\vskip -2\baselineskip plus -1fil

\begin{IEEEbiography}[{\includegraphics[width=1in,height=1.25in,clip,keepaspectratio]{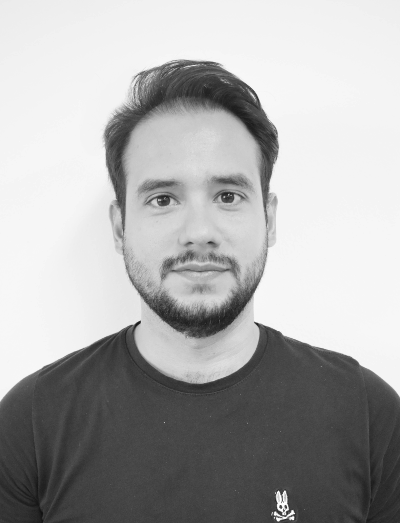}}]{Luis E. Rueda G.} received his bachelor’s degree in electronic engineering (cum laude) from the Universidad Industrial de Santander (UIS), in 2010, and his MSc. degree in electrical engineering from the Delft University of Technology in 2014. From 2010 to 2012, he was part of the IC Brazil Cadence program. As part of the program, he did an internship as an analog IC designer at Freescale Semiconductors from 2011 to 2012, working on the embedded memory group, where he co-authored a patent on a voltage ramp-up protection circuit. At TU Delft, he joined the Electronic Instrumentation Laboratory at TU Delft in September 2013, to conclude his master thesis in CO2 sensors, in collaboration with NXP Semiconductors. From December 2014 to September 2015, he worked as analog IC designer (TMC contractor) at IMEC, Belgium. He joined the Integrated Systems Research Group - OnChip at UIS in 2016, to start his Ph.D. research on machine learning (mainly deep learning) accelerators, while working as an adjunct professor at the electrical engineering school.
\end{IEEEbiography}

\vskip -2\baselineskip plus -1fil

\begin{IEEEbiography}[{\includegraphics[width=1in,height=1.25in,clip,keepaspectratio]{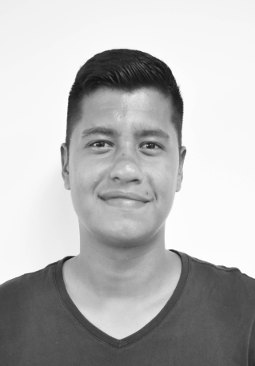}}]{Edward Silva} is a senior in the electronic engineering program at the Universidad Industrial de Santander (UIS) at Bucaramanga, Colombia. In 2017 he worked as an Electrical Circuit course tutor. Currently, he is a teacher assistant of Fundamentals of Analog Circuit course. His research interests include the implementation of analog circuits in FPGAs, in particular the design of digital low drop-out regulators.
\end{IEEEbiography}

\vskip -2\baselineskip plus -1fil

\begin{IEEEbiography}[{\includegraphics[width=1in,height=1.25in,clip,keepaspectratio]{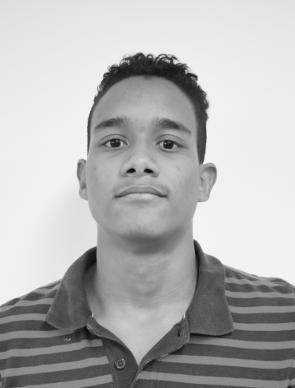}}]{Andres Centeno} is a senior in the electronic engineering program at the Universidad Industrial de Santander (UIS) at Bucaramanga, Colombia. In 2017 he worked as an Introduction to Engineering course tutor and in 2019 for the Electronic Devices course. His research interests include the implementation of analog circuits in FPGAs, in particular digital low drop-out regulators. 
\end{IEEEbiography}

\vskip -2\baselineskip plus -1fil

\begin{IEEEbiography}[{\includegraphics[width=1in,height=1.25in,clip,keepaspectratio]{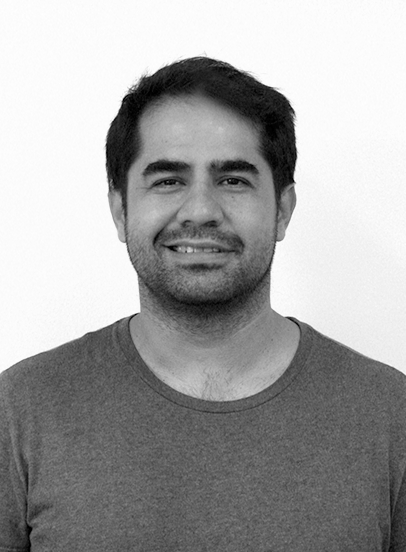}}]{Elkim Roa}
received the B.S.E.E. degree from Universidad Industrial de Santander at Bucaramanga, Colombia, in 1999, the M.S.E.E. degree from the University of Sao Paulo in 2003, Sao Paulo, Brasil, and the Ph.D. degree in electrical engineering from Purdue University in 2014 where he was a Fulbright scholar. From 2014 to 2016, Elkim was with Rambus Inc. where he was engaged in high-speed SERDES front-end design. Currently, he is an associate professor at Universidad Industrial de Santander, Colombia. His research interests include circuits and VLSI design for security, front-end circuits for high-speed interfaces and low-energy and efficient computing. He has authored and co-authored over 30 conference/journal publications. He holds two patents and has five pending patents in the area of integrated circuits. Dr. Roa had served on several technical program committees for IEEE CAS and IEEE SSCS conferences, and as a reviewer of IEEE Transactions on Microwave Theory and Techniques, IEEE Journal of Solid-State Circuits, and IEEE Transactions on Circuits and Systems TCAS-I. He currently serves as a technical program committee member of the IEEE International Symposium on Circuits and Systems (ISCAS) and the IEEE Custom Integrated Circuits Conference (CICC).
\end{IEEEbiography}

\end{document}